\documentclass[aps,groupedaddress,superscriptaddress,preprint]{revtex4}
\usepackage[dvipdfmx]{graphicx}
\usepackage[dvipdfmx]{epsfig}
\usepackage{color}
\usepackage{amsmath}
\usepackage{dcolumn}
\usepackage{bm}

\begin{document}
\title{Anomalous Nernst Effect in Nonmagnetic Nodal Line Semimetal PbTaSe$_2$}

\author{K. Yokoi$^*$}
\affiliation{Department of Physics, Osaka University, Toyonaka, Osaka 560-0043, Japan}

\author{H. Murakawa$^*$}
\affiliation{Department of Physics, Osaka University, Toyonaka, Osaka 560-0043, Japan}

\author{H. Sakai}
\affiliation{Department of Physics, Osaka University, Toyonaka, Osaka 560-0043, Japan}

\author{N. Hanasaki}
\affiliation{Department of Physics, Osaka University, Toyonaka, Osaka 560-0043, Japan}

\begin{abstract}
{PbTaSe$_2$ is a unique topological material, in which the number of nodal lines is expected to change at the structural transition between the lower temperature/pressure ``L'' phase and the higher temperature/pressure ``H'' phase. 
We report the anomalous Nernst effect attributed to the Berry curvature of nodal lines and its change with the structural transition. 
In the L phase, the Nernst coefficient ($S_{yx}$) shows the step-like magnetic field dependence reminiscent of the anomalous Nernst effect of nonmagnetic Dirac/Weyl semimetals. 
By applying hydrostatic pressure, we discovered that the amplitude of the anomalous component significantly decreases at the transition to the H phase, which might correspond to the partial annihilation of nodal line structures.
}
\end{abstract}

\maketitle

\begin{figure}
	\begin{center}
		\includegraphics[width=0.85\linewidth]{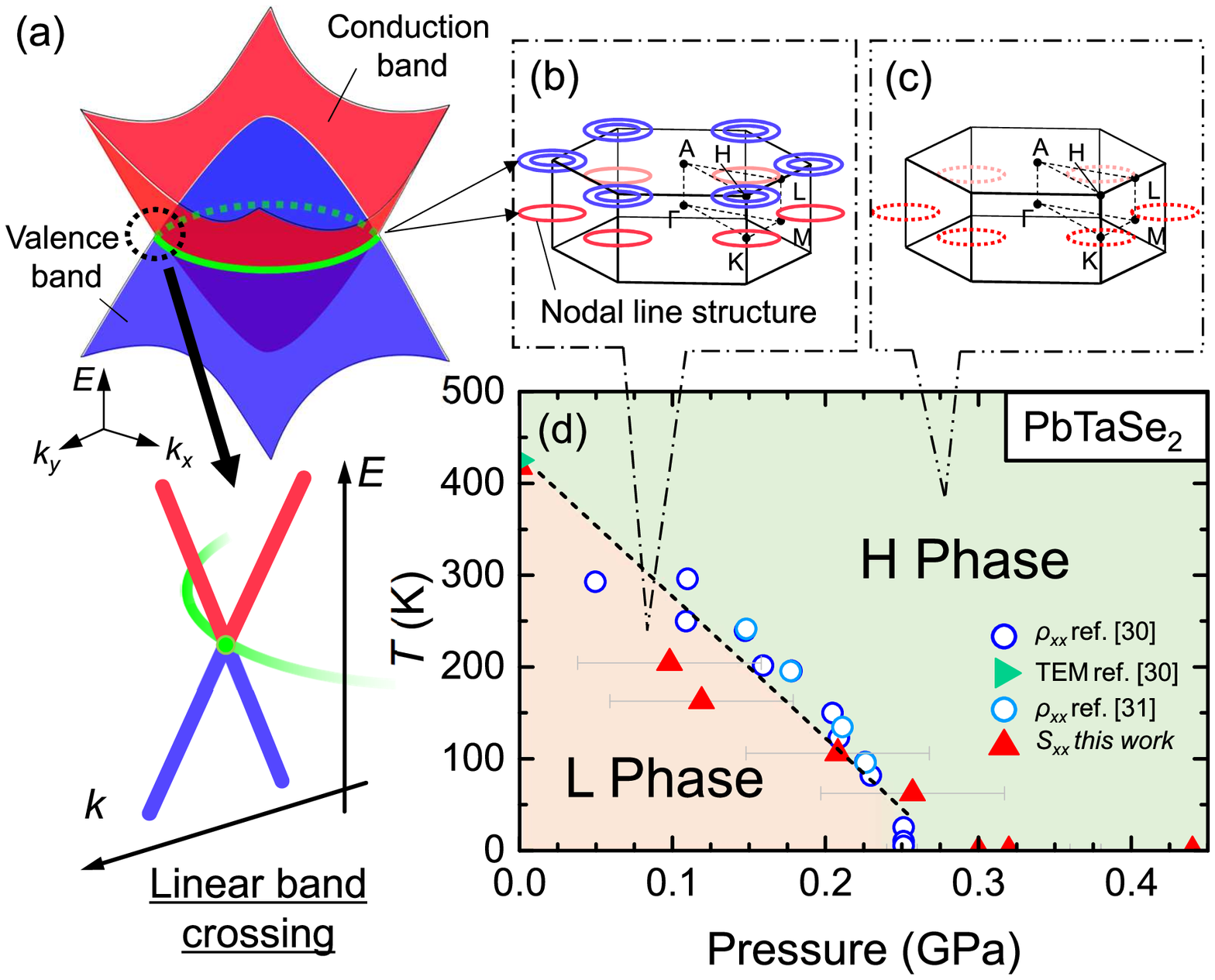}
		\caption[Structure]{\label{Fig1}(Color online) (a) Schematics of nodal line structure (green circle). The crossing of linear band dispersion occurs on the loop or line in the $k$-space, which is protected by the mirror or glide symmetry. (b) and (c) The Brillouin zone of PbTaSe$_2$. The red (on the $k_z = 0$ plane) and blue (on the $k_z = \pi / 2$ plane) circles represent the nodal lines. The dashed rings in (c) indicate the partially open bandgap. (d) Temperature versus pressure phase diagram of the crystal structure of PbTaSe$_2$. The open circles, the green triangle, and red triangles correspond to the transition temperatures in $\rho_{xx}$, high-temperature X-ray diffraction and transmission electron microscopy in refs.\cite{PRB224508, PRB064528}, and our Seebeck measurements, respectively.}
	\end{center}
\end{figure}

Dirac/Weyl fermions, the quasi-particle corresponding to the linear energy dispersion with a degenerated point (Dirac and Weyl cone), discovered in materials, such as the surface of topological insulators and graphene, have led to the exotic quantum phenomena. 
Currently, the search for new three-dimensional (3D) topological materials with those fermions and the study of their unique properties are the central subjects in condensed matter physics \cite{PRB195320,NatMat667,NatPhy550,PRB205101,PRL186806,PRX001029,PRB045135}.
The latest development of the study is the discovery of nodal line semimetal as a new type of 3D topological material that has the linear band crossing along the line or loop in the momentum space (Fig. \ref{Fig1} (a)) \cite{PRB235126,PRB081201,CPB117106,NatCom11696,JPSJ013708,NatPhy667,PRB205132}.
Characteristic phenomena reflecting the high-dimensional degenerated region \cite{PRB075138,PRB161105,PRL196603,PRB205132,NatCom14022} and the Berry curvature \cite{JPSJ013708,JPSJ093704,PRL026603,PMP1959,PRB16113} are expected in nodal line semimetals.

PbTaSe$_2$ is a nonmagnetic nodal line semimetal that has the potential to show exotic properties. 
This material is composed of transition metal dichalcogenide TaSe$_2$ and Pb atoms, (see the supplemental material Fig. S1) \cite{PhysB121}.
The band structure of PbTaSe$_2$ includes three types of ring-shaped nodal lines protected by a mirror symmetry on the $ab$-plane (Fig. \ref{Fig1} (b)), which arises from the band inversion between Ta-$5d$ and Pb-$6p$ bands around the K and H points in the Brillouin zone (Fig. S2 (a)) \cite{PRB020505,PRB165148,NatCom10556}.
Since PbTaSe$_2$ has the noncentrosymmetric crystal structure whose space group is represented as $P$\={6}$m2$, the spin degeneracy of those nodal rings is lifted to form the so-called Weyl loops.
Some experiments on PbTaSe$_2$ discuss the unique phenomena associated with those nodal line structures.
The Drum-head surface states are confirmed by angle-resolve photoemission spectroscopy \cite{NatCom10556} and quasi-particle interference \cite{SciAd1600894}.
The Seebeck and magnetic torque measurements, which detected the quantum oscillation and existence of the Berry phase in the nodal line structure, were discussed \cite{PRB104516}.
However, the contribution of nodal line fermions to transport phenomena is generally hard to distinguish from that of normal charge carriers.

Furthermore, PbTaSe$_2$ is the unique system expected to change the number of nodal line structures at the structural transition. 
Figure \ref{Fig1} (d) is the pressure vs. temperature phase diagram of PbTaSe$_2$.
The structural transition has been detected in x-ray diffraction (XRD) and transmission electron microscopy (TEM) measurements in high-temperature regions and resistivity measurement in hydrostatic pressure \cite{JSNM3173, PRB224508, PRB064528}.
``L phase''  and ``H phase'' denote the low-temperature/pressure phase and the high-temperature/pressure phase, respectively.
Theoretical calculation predicted that in the crystal structure of the H phase, Pb atoms change their position from the $1a$ Wyckoff site $(0,0,0)$ to the $1e$ Wyckoff site $(2/3,1/3,0)$, consequently decreasing the volume of the unit cell, (Fig. S1)  \cite{PRB224508, PRB064528}. 
This is consistent with the change in the lattice constant observed on XRD and TEM results \cite{PRB224508}.
The DFT calculation for the H phase based on the above crystal structure suggested that the two types of nodal lines represented as blue rings in Fig. \ref{Fig1} (b) disappear, and the red rings partially gap out to break the ring shape (Fig. \ref{Fig1} (c)) \cite{PRB064528}.
Thus, we can examine the influence of the topological electronic state change on the bulk transport phenomena.

In this study, we examine the Seebeck and Nernst effects of PbTaSe$_2$ in the L and H phases by applying hydrostatic pressure to catch the unique phenomena originating from the Berry curvature and change in topological electronic state.
We observed the step-like magnetic field dependence in the L phase, reminiscent of the anomalous Nernst effect in Dirac/Weyl semimetals reported in Refs. \cite{PRL136601,PRB201107,PRL196602,Nano6501,PRB115201}.
Furthermore, the temperature dependence of the Seebeck and Nernst coefficients shows the nontrivial peak structure probably associated with the linear band dispersion in the L phase (below $0.3$ GPa). 
However, in the H phase, we found that both the Seebeck and Nernst effects drastically altered their characteristics.
The anomalous term of the Nernst effect is significantly reduced in the H phase, consistent with the theoretically predicted partial annihilation of nodal line structures.
Our thermoelectric measurements capture the contribution of the Berry curvature to transport phenomena and a signature of the change of topological band structure.

Single crystals of PbTaSe$_2$ were synthesized by chemical vapor transport method, and plate-like crystals, whose typical dimension is approximately $3 \times 3 \times 0.05$ mm, were obtained (Fig. S3 (a)).
The sample quality was checked from the XRD spectrum, residual resistivity ratio (RRR), and superconducting transition temperature, (see the supplemental material for more detailed information). 
The measurement of thermoelectric properties was performed by the one-heater two-thermocouple methods, using the Physical Properties Measurement System (Quantum Design). 
The Nernst coefficient was measured in the vortex configuration \cite{PRB024510}.
For the thermoelectric measurement under hydrostatic pressure, the system shown in the inset of Fig. \ref{Fig3} (a) was built in the piston-type pressure cell, (for more details on the setup, see Fig. S3 (c)). 
The actual sample pressure was estimated by the Pb superconducting transition temperature.

\begin{figure}
	\begin{center}
		\includegraphics[width=1.05\linewidth]{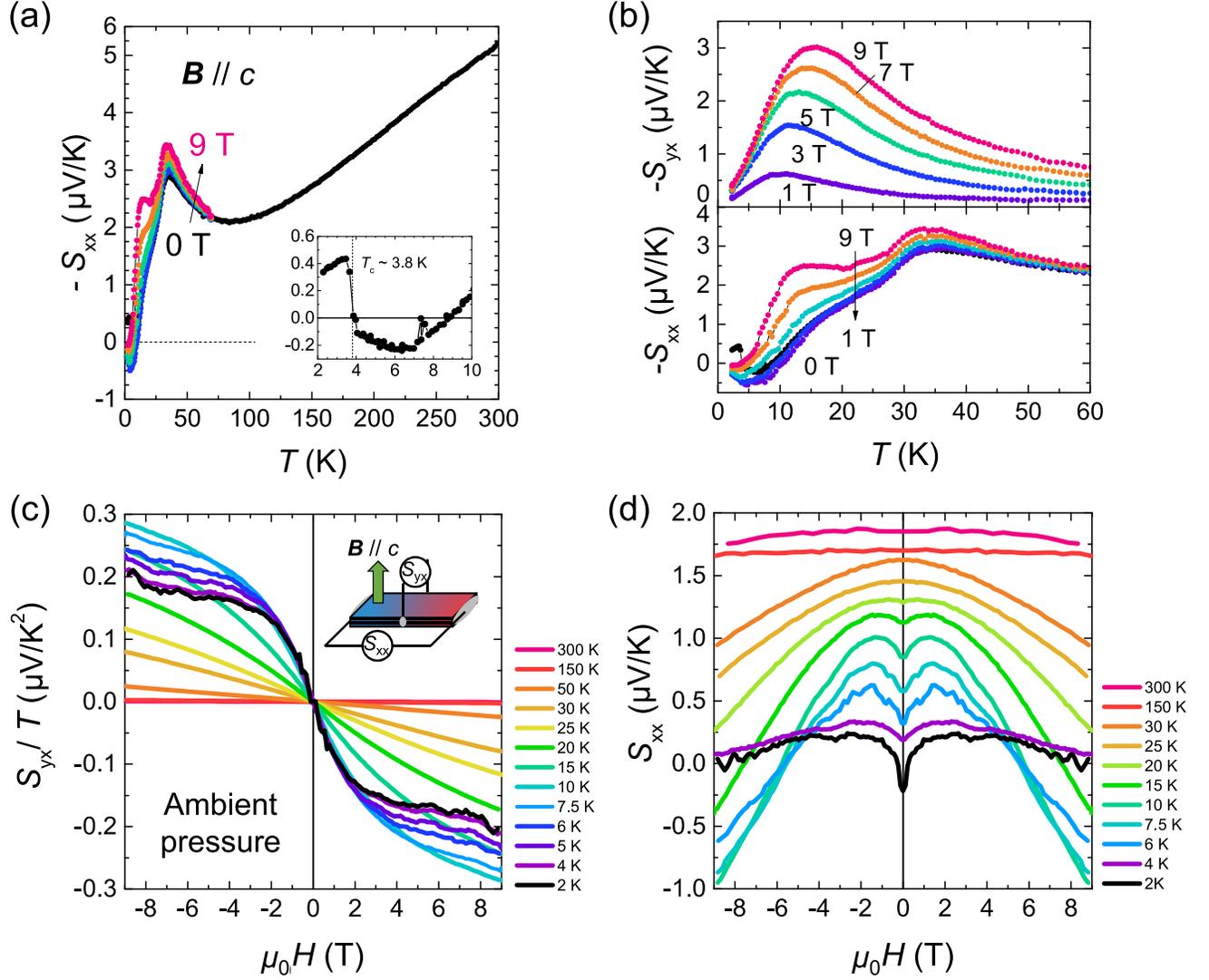} 
		\caption[Structure]{\label{Fig2}(Color online) Thermoelectric properties in PbTaSe$_2$ at ambient pressure. (a) Temperature dependence of $-S_{xx}$. The inset shows $-S_{xx}$ in the low-temperature range, in which the superconducting transition is evidently observed. (b) Temperature dependence of $-S_{yx}$ (upper panel) and $-S_{xx}$ (lower panel) in the magnetic fields (${\it B} \parallel c$ ). (c) and (d) Magnetic field dependence of $S_{yx}/T$ and $S_{xx}$ (with offsets about $0.03 \sim 7.2$ $\mu$VK$^{-1}$ except for $2$ K), showing similar features reported in Dirac/Weyl semimetal such as the saturation of $S_{yx}/T$ in high magnetic fields and the dip of $S_{xx}$ at zero magnetic fields. The oscillation of $S_{xx}$ in a high magnetic field ($>6$ T) is a quantum oscillation (for more detail, see the Supplemental materials.)  }
	\end{center}
\end{figure}

\begin{figure}
	\begin{center}
		\includegraphics[width=1.0\linewidth]{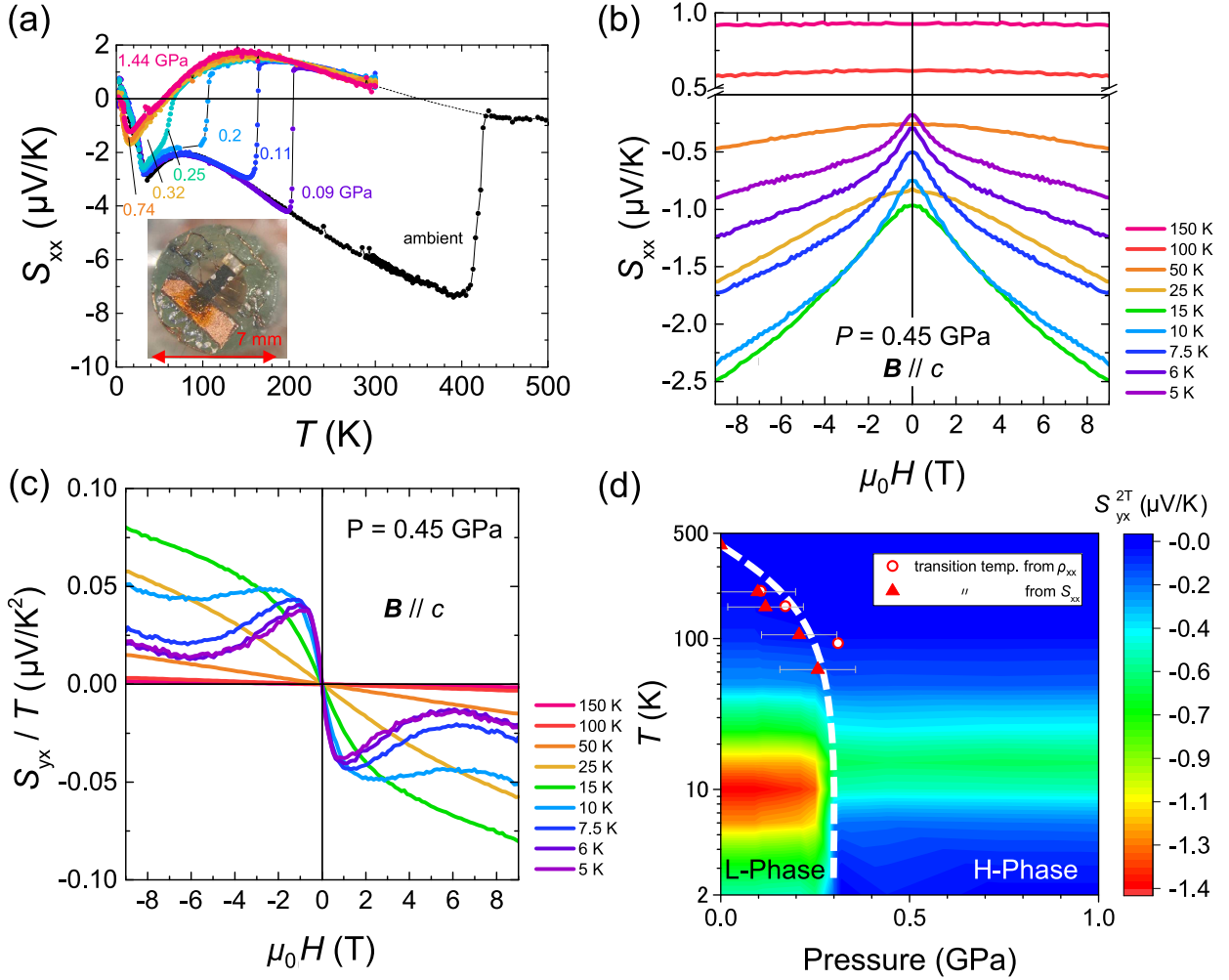} 
		\caption[Structure]{\label{Fig3}(Color online) Thermoelectric properties of PbTaSe$_2$ under pressure. (a) Temperature dependence of $S_{xx}$ under various pressures. As the pressure increases, the transition temperature shifts to the lower temperature range. Although thermal hysteresis is observed, only the cooling process is shown for simplicity. The inset is the picture of a thermoelectric measurement system in hydrostatic pressure. (b) and (c) Magnetic field dependence of $S_{xx}$ and $S_{yx}$ in high - pressure structure phase ($P = 0.45$ GPa). (d) Pressure-temperature phase diagram and color contour map of $S_{yx}^{2\text{ T}}$. The red triangles and open circles represent the transition points obtained by the thermoelectric and resistivity measurements, respectively. The dashed white curve is an eye guide.}
	\end{center}
\end{figure}

Figure \ref{Fig2} (a) shows the temperature dependence of the Seebeck coefficient ($S_{xx}$) at ambient pressure.
$S_{xx}$ is proportional to temperature between $300$ and $100$ K, consistent with the semiclassical Boltzmann model.
However, $S_{xx}$ shows the peak at 35 K and hump at 15 K, which cannot be explained by the simple Boltzmann model.
The peak at 35 K is insensitive to magnetic fields, while the hump at 15 K shows clear field dependence and develops its amplitude as increasing magnetic fields (the upper panel of Fig. \ref{Fig2} (b)). 
These structures were not observed in polycrystalline PbTaSe$_2$ \cite{JSNM3173}.
Furthermore, sign inversion of $S_{xx}$ was observed near $8$ K and $4$ K.
The former seems to reflect the multipolarity of carriers in PbTaSe$_2$, and the latter corresponds to superconducting (SC) transition, since the temperature at the center of the $S_{xx}$ jump coincides with the SC transition temperature in previous reports \cite{PRB020505,NatCom10556,JSNM3173,PRB224508,PRB064528,PRB020503,PRB060506,PRB184510,PRB214504}.
A nonzero $S_{xx}$ observed in the SC state probably originates from the gold and/or copper used in the leading wire and heat sink/diffuser.

Figure \ref{Fig2} (b) shows the expanded view of $S_{yx}$ and $S_{xx}$ in magnetic fields below 60 K. 
The peak structure in $S_{yx}$ appears close to the $S_{xx}$ hump at 15 K, which is sensitive to magnetic fields.
The theoretical simulation of thermoelectric coefficients for Weyl fermions suggests the appearance of the peak structure in $S_{yx}$, where the peak-top temperature in high ($\sim9$ T) and low ($\sim1$ T) magnetic fields have the following relationship \cite{PRB161404}: $T^{\rm peak}_{\rm low H} / T^{\rm peak}_{\rm high H} = \sqrt{3}/\pi \approx 0.55$, which is consistent with our experimental value, $T^{\rm peak}_{1 \rm T} / T^{\rm peak}_{9 \rm T} \approx 8.9$ K $/15.5$ K $\approx 0.57 \pm 0.06$.
Therefore, the $S_{yx}$ peak structure at $15$ K is considered to be attributed to the Weyl loops in PbTaSe$_2$.
Since the $S_{xx}$ hump is observed at the same temperature where the $S_{yx}$ peak appears, Weyl loops may also relate to its origin.
However, the origin of the peak seen in $S_{xx}$ at $35$ K remains unclear but possibly relates to the multicarrier and phonon drag effect.
Moreover, note that the $S_{yx}$ peak amplitude normalized by magnetic fields approximately $\sim-0.33\;\mu \rm V \rm K ^{-1} \rm T^{-1}$ is notable among trivial metals and comparable to novel metals showing a superior Nernst effect such as charge density wave compound NbSe$_2$ ($\sim-0.12\;\mu \rm V\rm K^{-1} \rm T^{-1}$) and heavy fermion compound CeRu$_2$Si$_2$ ($\sim-0.16\;\mu \rm V\rm K^{-1} \rm T^{-1}$) \cite{JPCM113101}.

The magnetic field dependence of $S_{yx}$ and $S_{xx}$ are shown in Figs. \ref{Fig2} (c) and (d), respectively.
In $S_{yx}$, quasi-linear magnetic field dependence is observed at high temperatures, and step-like behavior reminiscent of the anomalous Nernst effect appears below $15$ K.
Furthermore $S_{xx}$ also shows the unique magnetic field dependence which has a dip at zero magnetic fields in the low temperature range.
Note that those dip does not correspond to the SC transition because it appears above $T_c$. 
Those unique magnetic field dependences of $S_{yx}$ and $S_{xx}$ in PbTaSe$_2$ are reminiscent of Dirac semimetal Cd$_{3}$As$_{2}$ and Weyl semimetal TaX (X = P, As) that show the anomalous Nernst effect reflecting the Berry curvature \cite{PRL136601,PRB201107}.
Furthermore, we measured $S_{yx}$ up to 15 T and confirmed that the step-like behavior consistent with the anomalous Nernst effect is preserved (Fig. S5). 
Above 5 T, we observed clear quantum oscillation in $S_{xx}$ and $S_{yx}$.
From their frequencies obtained by the FFT analysis, it was observed that the Fermi level of our sample was nearly the same as in the previous report \cite{PRB104516} and close to the nodal rings. 
(For more detailed information, see Supplemental material.)

Next, we discuss the influence of hydrostatic pressure on thermoelectric properties.
Figure \ref{Fig3} (a) shows the temperature dependence of the Seebeck coefficient $S_{xx}$ at various pressures.
Sharp jumps of $S_{xx}$ occur at the structural transition temperature and shift to lower temperatures with increasing pressure, which is consistent with previous resistivity data \cite{JSNM3173, PRB224508, PRB064528}.
Although evident thermal hysteresis is observed in $S_{xx}$, for simplicity, we show only the cooling curve.
Figure \ref{Fig3} (b) shows the magnetic field dependence of $S_{xx}$ at $0.45$ GPa after the structural transition.
The amplitude and behavior of the magnetic field dependence of $S_{xx}$ are almost independent of pressure.
(The data for the other pressure is shown in the Supplemental material Section V).

\begin{figure}
	\begin{center}
		\includegraphics[width=0.9\linewidth]{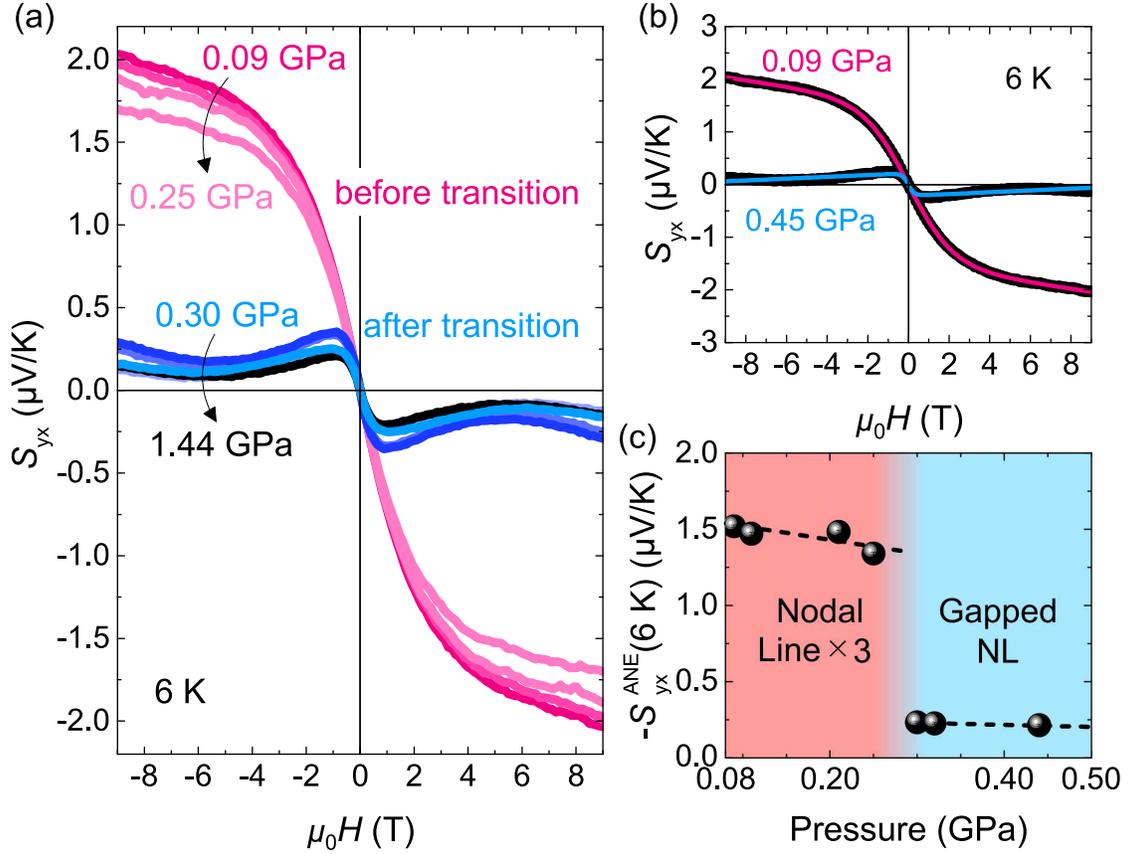} 
		\caption[Structure]{\label{Fig4}(Color online) (a) Pressure dependence of $S_{yx}$ at $6$ K, where step-like behavior is prominent. The warm or cold colored curves represent the data in pressure below or above structural-transition pressure, respectively. (b) Fitting analysis results for 0.09 GPa (below the transition pressure) and 0.45 GPa (above the transition pressure). The bold black and colored curves represent the experimental data and simulations, respectively. (c) Pressure dependence of the anomalous Nernst term $S_{yx}^{ANE}$ at 6 K. The anomalous component abruptly drops by a factor of $1/7$, consistent with the density reduction of the nodal line fermions.}
	\end{center}
\end{figure}

Figure \ref{Fig3} (c) shows the magnetic field dependence of the Nernst coefficient $S_{yx}/T$ in the H phase ($P=0.45$ GPa).
The quasi-linear magnetic field dependence of $S_{yx}$ is observed at high temperatures, and the step-like feature gradually develops with cooling.
We consider that such a magnetic field dependence is attributed to the anomalous Nernst effect originating from the remaining nodal line structures predicted by the theory  \cite{PRB064528}. 
However, the field dependence of $S_{yx}$ is slightly different from that in the L phase.
The $S_{yx}$ amplitude takes a local minimum near 5 T below 15 K, which may be related to the actual shape of the band dispersion of the remaining nodal region that is neither perfect linear nor parabolic.
Figure \ref{Fig3} (d) shows the color contour plot of the Nernst coefficient at 2 T, in which the structural transition temperatures are determined by resistivity and Seebeck measurements and indicated by the open circles and triangles, respectively. 
Note that the heat map feature is almost independent of field (below 9 T).
In a high-temperature region above 100 K, the amplitude of the Nernst signal in the L phase is similar to that in the H phase, and insensitive to pressure within the identical phase.
However, below 100 K, the $S_{yx}$ amplitude dramatically changes at the transition between the L and H phases.
Figure \ref{Fig4} (a) shows the magnetic field dependence of $S_{yx}$ at 6 K in both L and H phases.
The $S_{yx}$ amplitude decreases by a factor $\sim1/7$ after the transition to the H phase.
If the step-like magnetic field dependence mainly comes from the anomalous Nernst effect driven by the Berry curvature, the dramatic change of $S_{yx}$ between the L and H phases might reflect the change in the number of nodal line structures suggested by theoretical calculation \cite{PRB064528}.

To evaluate the anomalous Nernst effect in both phases, we derive the anomalous term by fitting $S_{yx}$ into a simple formula:
\begin{equation} 
\label{simple Syx}
S_{yx} = S_{yx}^{ANE} \times \tanh{\frac{B}{B_s}} + S_{yx}^{Linear}\times B,
\end{equation}
where the first term represents the anomalous Nernst effect \cite{PRL136601,PRB201107,PRL196602}.
$S_{yx}^{ANE}$ and $S_{yx}^{Linear}$ are the corresponding amplitude of each term, and $B_s$ is the saturation field of the ANE term.
We assume that the second background term is proportional to the magnetic field as a first approximation (For more detailed information, see Supplemental material). 
Figure \ref{Fig4} (b) shows the fitting results at 6 K, where the field profiles of $S_{yx}$ for both phase are well reproduced by Eq.\ref{simple Syx}.
Figure \ref{Fig4} (c) shows the pressure dependence of the amplitude on the anomalous Nernst term.
We observed that the amplitude of $S_{yx}^{ANE}$ in the H phase is less than $\sim 1/7$ of that in the L phase, which is consistent with the theoretical estimation that the density of the nodal line fermion in the H phase is less than one-third of that in the L phase \cite{PRB064528}.

In conclusion, we studied the thermoelectric properties in the nodal line semimetal PbTaSe$_2$ and investigated how the Seebeck and Nernst effects change upon the pressure-induced structural transition accompanied by the change in the number of nodal lines.
We discovered that the Nernst effect is quite sensitive to the number of nodal lines. 
Our results exhibit the contribution of the nodal line fermions to the transport phenomena that was not achieved in previous studies.

\section*{Aknowledgements}
This work is partially supported by the Sasakawa Scientific Research Grant from the Japan Science Society, Asahi Glass Foundation, JST PRESTO (No. JPMJPR16R2), and JPSJ KAKENHI (Grant No. 18J04226, 19H01851,20J11036, 21H00147). 
\section*{$*$Corresponding author}
yokoi@gmr.phys.sci.osaka-u.ac.jp,
murakawa@phys.sci.osaka-u.ac.jp

\newpage
\newpage
\newpage
\newpage

\end{document}


\makeatletter
\renewcommand{\ref}[1]{S#1}
\renewcommand{\cite}[1]{[S#1]}
\renewcommand{\@biblabel}[1]{[S#1]} 
\makeatother
\renewcommand{\thefigure}{S\arabic{figure}}
\renewcommand{\thetable}{S\arabic{table}}
\renewcommand{\theequation}{S\arabic{equation}}

\title{Supplemental Material: Anomalous Nernst Effect in Nonmagnetic Nodal Line Semimetal PbTaSe$_2$}

\author{K. Yokoi$^*$}
\affiliation{Department of Physics, Osaka University, Toyonaka, Osaka 560-0043, Japan}
\author{H. Murakawa$^*$}
\affiliation{Department of Physics, Osaka University, Toyonaka, Osaka 560-0043, Japan}
\author{H. Sakai}
\affiliation{Department of Physics, Osaka University, Toyonaka, Osaka 560-0043, Japan}
\author{N. Hanasaki}
\affiliation{Department of Physics, Osaka University, Toyonaka, Osaka 560-0043, Japan}
\maketitle
\section*{$*$Corresponding author}
yokoi@gmr.phys.sci.osaka-u.ac.jp, murakawa@phys.sci.osaka-u.ac.jp

\section{Crystal structure of the high-temperature/pressure phase}
Figure \ref{1} shows the crystal structure in the L and H phases of PbTaSe$_2$, theoretically indicated in previous reports \cite{1,2}.
The Pb layers slide to the TaSe$_2$ layer, and Pb atoms move to the center of the Se triangular prism through the structural transition.
Since the basic structure of each layer is preserved, the structural transition does not change the configuration of the unit cell and space group.

\begin{figure}[b]
	\begin{center}
		\includegraphics[width=1.0\linewidth]{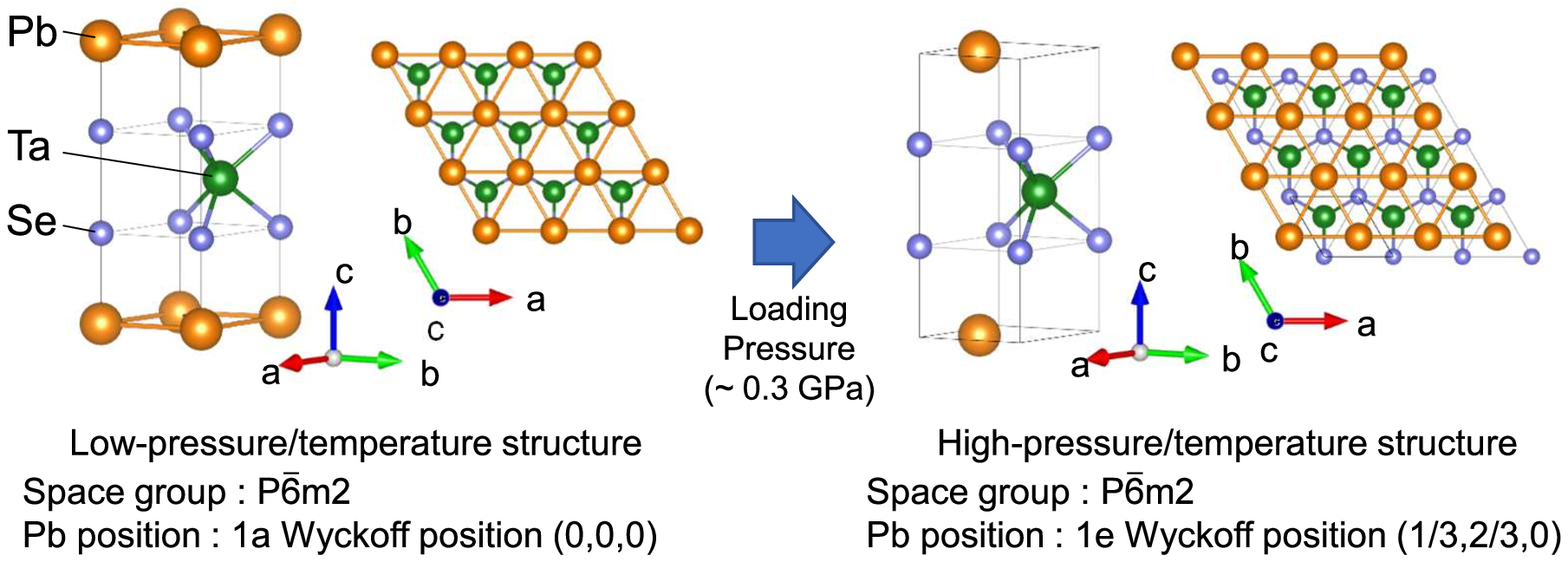}
		\caption[Structure]{\label{1}(Color online)  Crystal structure of PbTaSe$_2$ in low-temperature/pressure phase (left) and high-temperature/pressure phase (right) \cite{3}. }
	\end{center}
\end{figure} 
\clearpage
\section{electronic structure of PbTaSe$_2$}
We used the WIEN2k package \cite{4} to calculate the band structure and Fermi surfaces of PbTaSe$_2$.
These calculations were performed with GGA-PBE exchange-correlation potential \cite{5} and full-potential LAPW (FLAPW) basis set, considering the spin-orbit interaction.
Figures \ref{2} (a)-(c) show these results.
The bands near the $\Gamma$ point are dominated by Ta-5$d$ orbits, which form 3D hole pockets at the center of the Brillouin zone (Fig. \ref{2} (b)). 
The quasi-2D hexagonal Fermi surfaces surrounding the hole pockets consist of both Ta- and Pb-orbits.
These Fermi surfaces provide trivial carriers.
The inversion between Pb-conduction and Ta-valence bands occurs around the K and H points, forming the nodal line structures.
Thus, the cylindrical Fermi surfaces around the K and H points accommodate the nodal line fermions.

\begin{figure}[b]
	\begin{center}
		\includegraphics[width=1.0\linewidth]{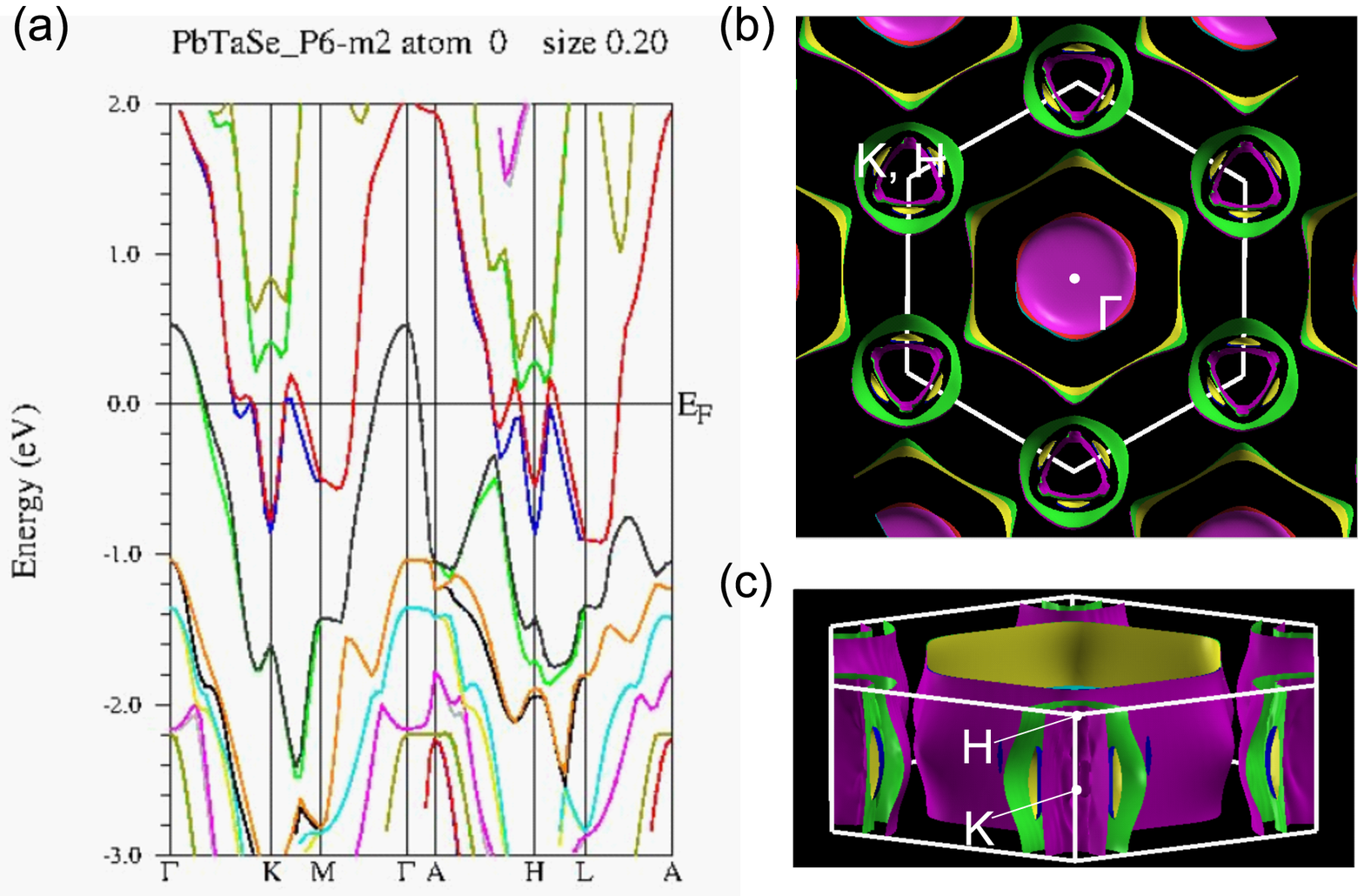}
		\caption[Structure]{\label{2}(Color online) (a) Band structure of PbTaSe$_2$ with spin-orbit interaction. (b) and (c) Fermi surfaces in PbTaSe$_2$.}
	\end{center}
\end{figure} 
\clearpage
\section{Experimental methods}
\subsection{Crystal growth}
The stoichiometric amounts of Pb ($99.99\%$, powder), Ta ($99.99\%$, powder) and Se ($99.999\%$, shot) were sealed into a silica tube ($11$ and $13$ mm inner and outer diameters, respectively) under approximately $10^{-4}$ Pa vacuum. 
The tube was heated to $950^\circ\mathrm{C}$ in a box furnace for three days to prevent the explosion of the silica tube by evaporation of unreacted selenides.
The obtained polycrystalline PbTaSe$_2$ and transport agent PbCl$_2$ were mixed and put into a silica ampule ($20$ and $23$ mm inner and outer diameters, respectively) under the N$_2$ gas atmosphere.

The PbCl$_2$ amount was adjusted below $2$ mg/cm$^3$ to prevent the ampule from exploding.
The ampule was sealed under approximately $10^{-4}$ Pa vacuum for the chemical vapor transport method and heated in the $3$-zone furnace at $900^\circ\mathrm{C}$ for the source side and $800^\circ\mathrm{C}$ for the sink side for three weeks.
Subsequently, thick plate-like soft and flexible single crystals with metallic luster were obtained (Fig.\ref{3}(a)).
We shaped the sample into a rectangular bar with a surgical knife.

\subsection{The thermoelectric measurement in ambient pressure}
The thermoelectric measurement in ambient pressure was performed with a homemade PPMS pack (Fig. \ref{3} (b)).
A sample was bridged between the heat sink Cu block and heat diffuser Cu block with resistive heaters.
The sample and Cu blocks were connected thermally and electrically by silver paste.
The leading wires were attached to the Cu blocks, and the Seebeck voltage was measured through them.
Based on the assumption that the temperature at the ends of the sample was the same as that of the Cu blocks, the temperature difference between those blocks measured using E-type thermocouples approximates the temperature difference of the sample.
The Nernst voltage was measured by the contacts on the side of the sample in the so-called vortex configuration.

\subsection{The thermoelectric measurement in the hydrostatic pressure}
\begin{figure}[b]
	\begin{center}
		\includegraphics[width=0.5\linewidth]{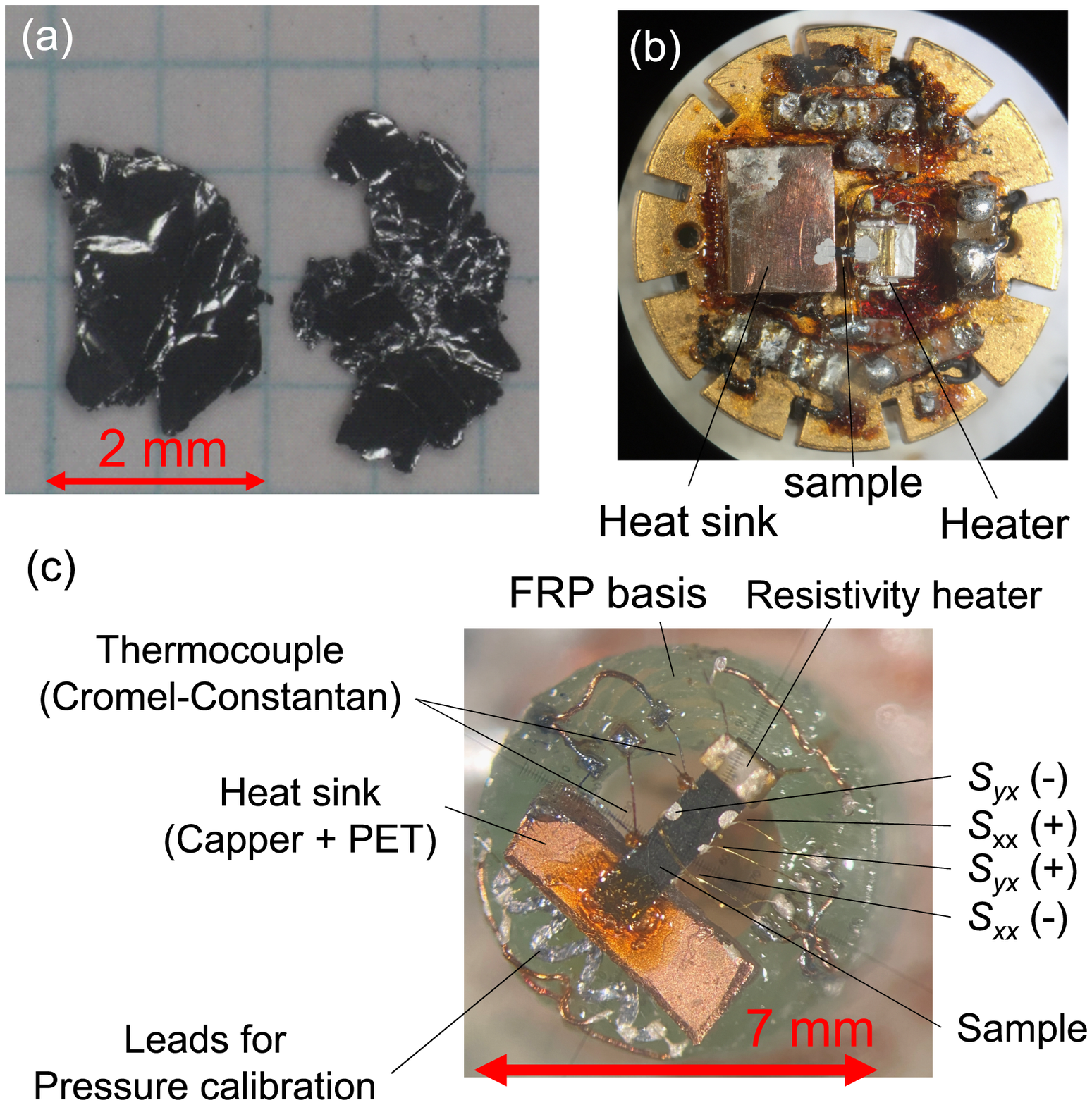}
		\caption[Structure]{\label{3}(Color online)  Photograph of (a) synthesized single crystals, (b) homemade measurement unit on the physical properties measurement system (PPMS) package for the thermoelectric measurement, and (c) details of the measurement system in the hydrostatic pressure.  }
	\end{center}
\end{figure}

Figure \ref{3}(c) shows the details of the thermoelectric measurement system in hydrostatic pressure.
The cylindrical fiber reinforced plastics basis was attached to a Cu plug to construct a measurement system on it.
The Cu-coated plastic plate was used for the heat sink, and it contacted the end of the sample thermally by varnish.
A resistive heater (1 k$\Omega$) was stuck on another end of the sample.
Then, the width of the sample was adjusted to be smaller than that of the heater to avoid thermal imbalance in the sample.
For measuring thermoelectric coefficients, we attached thermocouples and voltage terminals independently, that is, we employed the 4-probe method. 
The deviation of the Seebeck coefficient from that in the 2-probe method (ambient pressure measurement) due to the difference in position between thermocouples and terminals was corrected by a constant factor.
The actual pressure was calculated from the pressure dependence of the superconducting transition temperature of a small lead wire put together with the sample based on the empirical formula: $P_{eff} = (T_c(P)-7.15 ) / 0.384$.
We used Daphene7373 as the pressure medium.
\clearpage
\section{Sample characterization}
\subsection{x-ray diffraction measurement}
It is hard to perform single-crystal XRD owing to the unavoidable natural waving of as-grown crystals with film shape.
For determining the crystal structure, we performed diffraction measurement for the (00$l$) plane in an as-grown single crystal, using a powder XRD instrument and calculated the lattice constant for the $c$-axis.
Figure \ref{4}(a) shows the obtained (00$l$) spectrum and the inset shows the (001) diffraction.
Sharp peaks whose positions coincide with those of the simulation appeared.
The lattice constant was calculated to be $\approx 9.38$\AA, which is consistent with previous reports \cite{6}.

\subsection{resistivity measurement}
\begin{figure}[h]
	\begin{center}
		\includegraphics[width=1.0\linewidth]{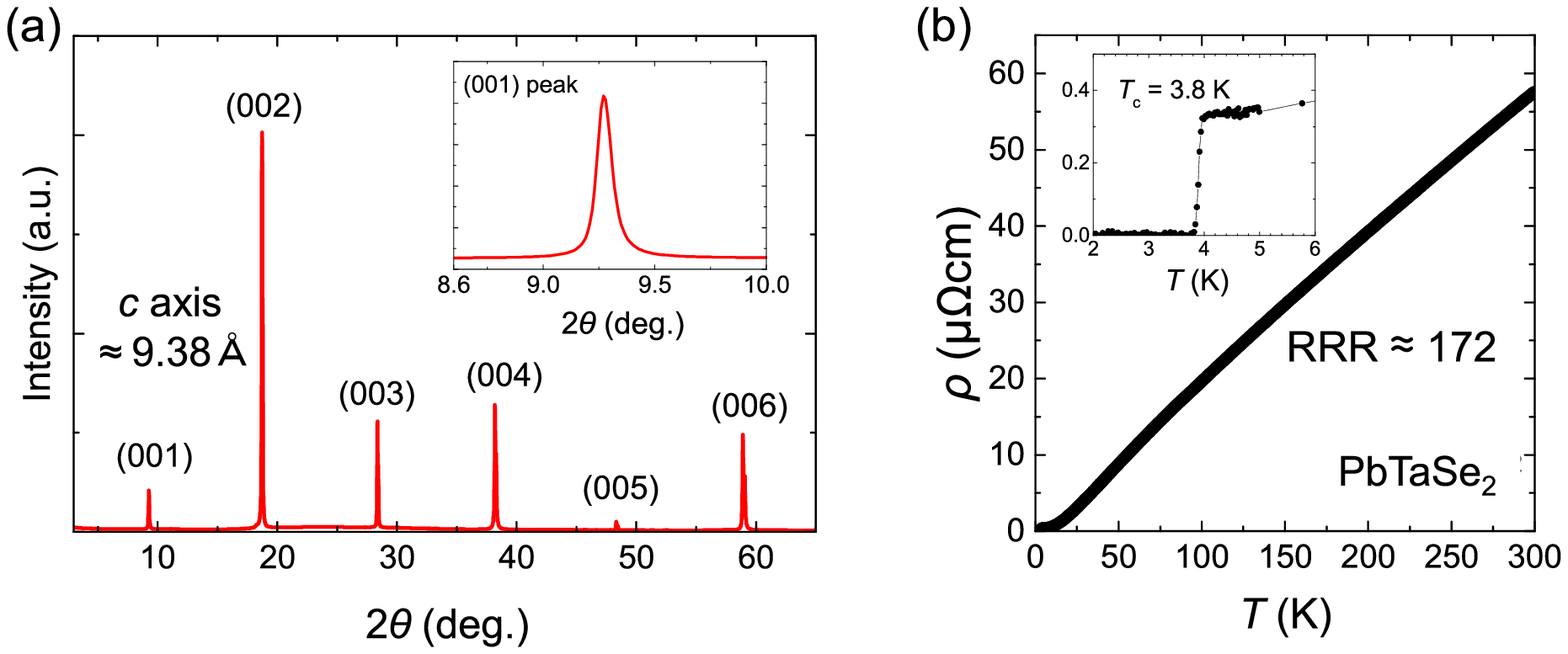}
		\caption[Structure]{\label{4}(Color online) 
			(a) X-ray diffraction spectrum for the ($00l$) plane. The inset is a magnified view around ($001$) peak. (b) Temperature dependence of resistivity. It shows a large residual resistivity ratio $\approx 172$ and a sharp superconducting transition at $3.8$ K (inset), indicating the high quality of our crystals. }
	\end{center}
\end{figure}

PbTaSe$_2$ is a superconductor with a superconducting transition temperature ($T_c$) $\approx 3.8$ K \cite{7}.
We checked the transition temperature of our sample in resistivity measurements for sample quality verification.
Figure \ref{4}(b) shows the temperature dependence of resistivity of our sample.
The sharp superconducting transition occurs at $3.8$ K, as shown in the inset, which is consistent with previous reports \cite{7}.
The residual resistivity ratio (RRR) is found to be $\approx 172$. 
\clearpage
\section{Thermoelectric properties and the quantum oscillation in high magnetic fields}
\begin{figure}[b]
	\begin{center}
		\includegraphics[width=1.0\linewidth]{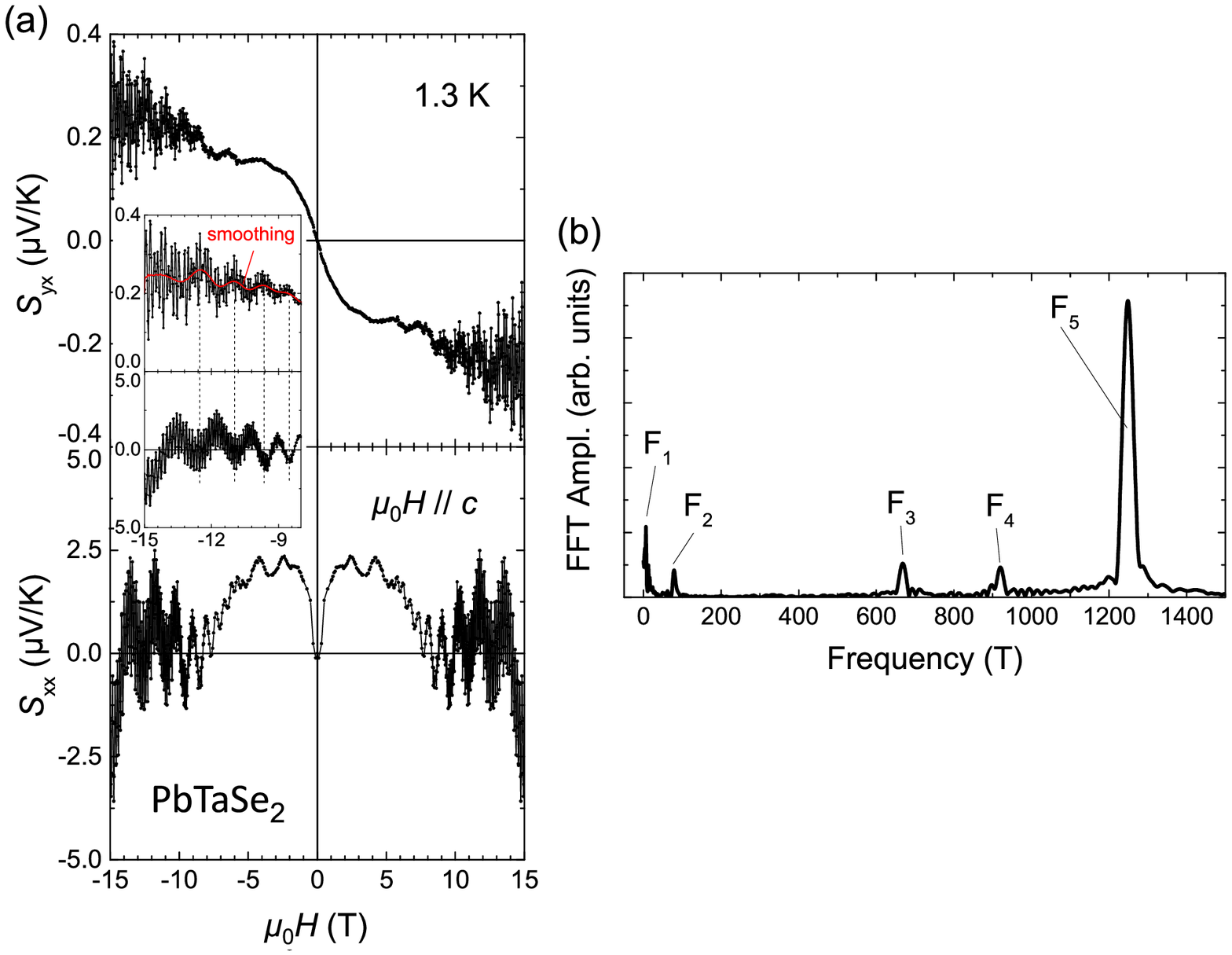}
		\caption[Structure]{\label{5}(Color online)  
			(a) Nernst and Seebeck effect up to 15 T in high magnetic field. The Nernst effect shows up to 15 T saturation behavior that is the characteristic feature of the anomalous Nernst effect. Quantum oscillations appear in $S_{yx}$ and $S_{xx}$; those on $S_{yx}$ are out of phase from $S_{xx}$ as shown in the inset. (b) Fast Fourier transform (FFT) spectrum for the quantum oscillation on $S_{xx}$. For each frequency, see Table S1. }
	\end{center}
\end{figure}

To investigate thermoelectric properties above 9 T, we measured them in higher magnetic fields up to 15 T using a superconducting magnet (Oxford Co. Ltd.).
Figure \ref{5} (a) illustrates the results of $S_{yx}$ and $S_{xx}$ measurements.
The Nernst coefficient $S_{yx}$ shows only the monotonical change with a constant slope up to 15 T, maintaining the characteristic step feature expected in the anomalous Nernst effect.

Furthermore, the quantum oscillation appears in both the Nernst and Seebeck effects.
$S_{xx}$ clearly shows high- and low-frequency oscillations.
$S_{yx}$ shows the similar oscillations, but their peaks and dips are slightly unclear, as shown in the inset of Figure \ref{5} (a). 
We subtracted the background by fitting the polynomial function to extract the oscillation and performed an fast Fourier transform (FFT) to analyze the frequency.
Figure \ref{5} (b) shows the FFT results for $S_{xx}$.
The five peaks appear in the spectrum, corresponding to five carrier pockets whose frequencies are summarized in Table S1.
The FFT spectrum obtained from $S_{yx}$ is almost the same as that of $S_{xx}$.
Our FFT results agree with previous magnetic torque and Seebeck Fermi level and electronic state in our sample are similar to those in the previous report \cite{8} .

\begin{table}[b]
	\begin{center}
		\caption[Structure]{Frequency of each component of quantum oscillation obtained from $S_{xx}$ in our sample (upper table) and previous report (lower table) \label{table2}}
		\begin{tabular*}{120mm}{@{\extracolsep{\fill}}cccccc}
			\hline\hline
			& $F_1$ & $F_2$ & $F_3$ & $F_4$ & $F_5$\\
		    \hline
			This work & 6.2 T & 79.3T & 677.4 T & 920.3 T & 1248.1 T\\
			Ref. \cite{8} & 5.8 T & 80.3 T & 672 T & 910 T & 1249 T\\
			\hline\hline
		\end{tabular*}
	\end{center}
\end{table} 
\clearpage
\clearpage

\section{The Nernst effect and Seebeck effect in hydrostatic pressure at each temperature}
\begin{figure}[b]
	\begin{center}
		\includegraphics[width=0.95\linewidth]{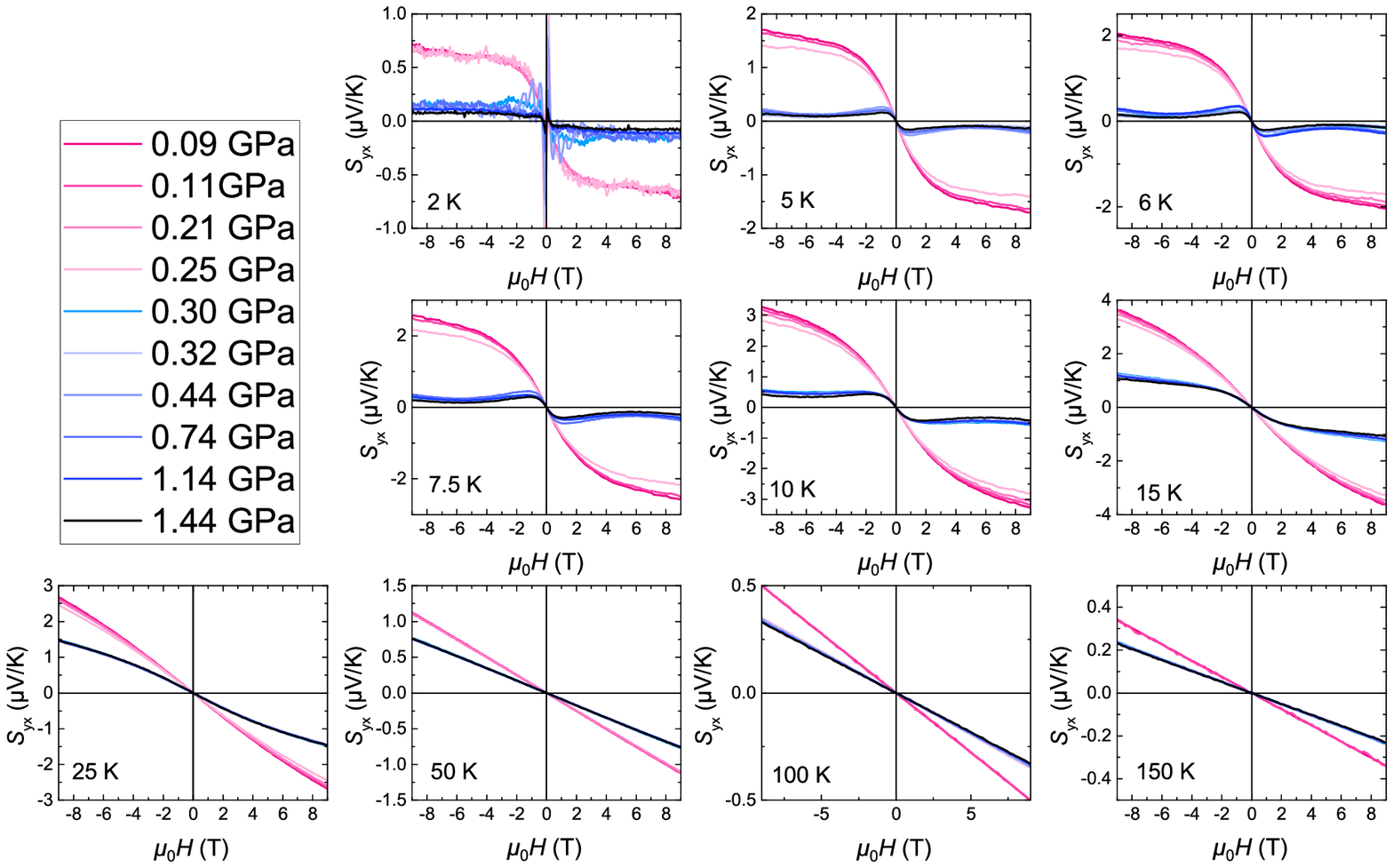}
		\caption[Structure]{\label{6}(Color online) Nernst effect in hydrostatic pressure at each temperature. The difference in the amplitude between the low- and high-pressure phase becomes evident with decreasing temperature. This probably indicates the change of the main factor in the Nernst effect from the thermally excited trivial carriers to the nodal line carriers. }
	\end{center}
\end{figure}

\begin{figure}[t]
	\begin{center}
		\includegraphics[width=1.0\linewidth]{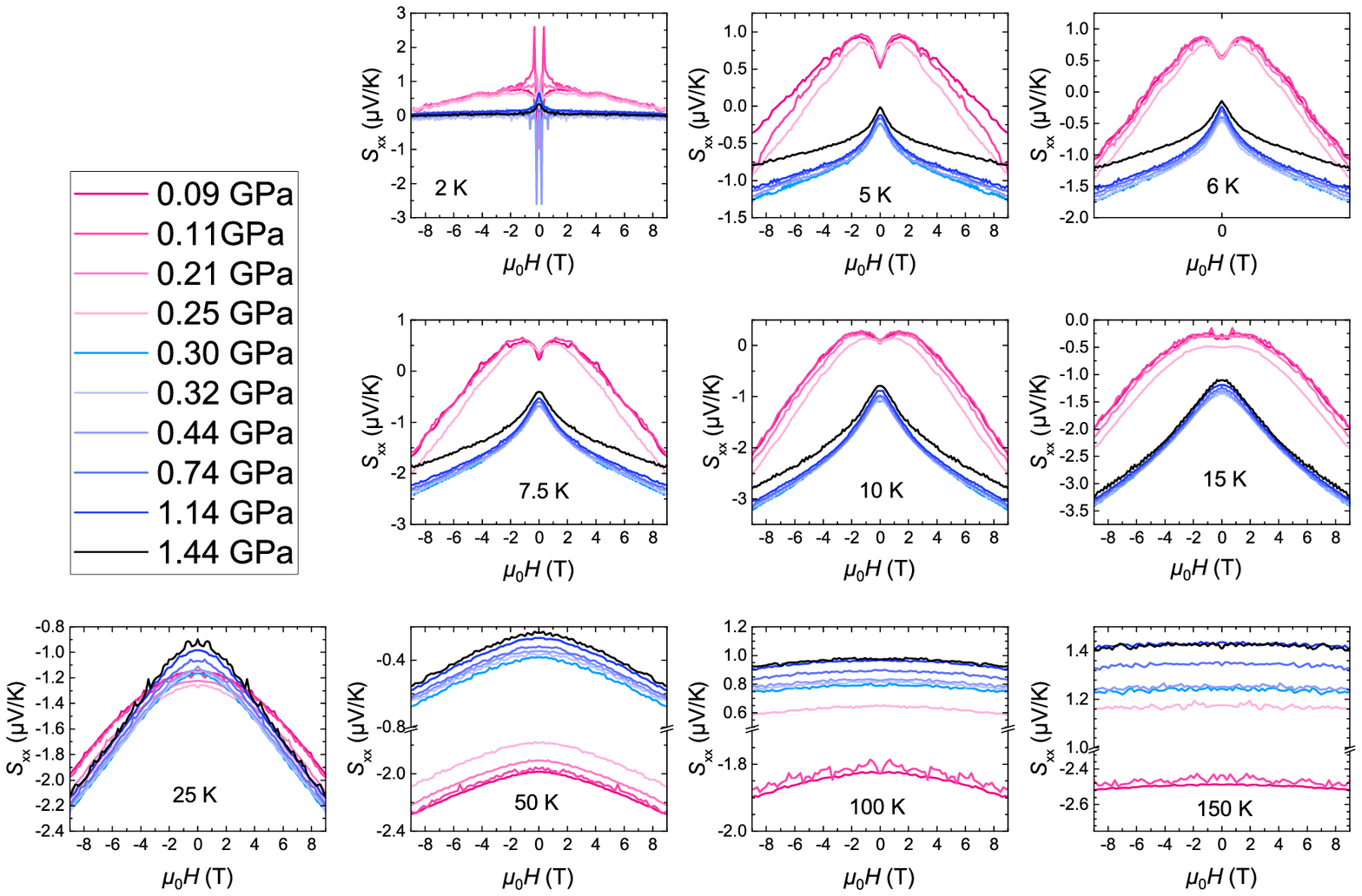}
		\caption[Structure]{\label{7}(Color online) Seebeck effect in hydrostatic pressure at each temperature. A sign change in the high-temperature range probably reflects the polarity of main carriers.}
	\end{center}
\end{figure}

Figures \ref{6} and \ref{7} show the Nernst and Seebeck coefficients, respectively, in the hydrostatic pressure at each temperature.
The difference in the Nernst coefficient $S_{yx}$ between low- and high-pressure phases becomes clear with cooling, and the ratio of the latter to the former reaches $\sim 1/7$ at 5 K.
The dramatic change of $S_{yx}$ with structural transition, clearly seen in the low-temperature range where thermal excitation of carriers is suppressed, indicates a significant variation in the electronic state.
The Seebeck coefficient $S_{xx}$ changes its sign above 50 K, suggesting the sign change of electrical charge of main carriers, while the Nernst’s sign remains unchanged.
Thus, we deduce that the sign of the Nernst coefficient does not directly reflect the sign of the charge carriers.

\clearpage

\section{Fitting the Nernst coefficient}
The anomalous Nernst effect originating from the Berry curvature is generally proportional to the function of $\tanh (B/B_{s})$ \cite{9,10}. 
Thus, we employ the same function to estimate the size of the anomalous Nernst effect.

However, the total Nernst signal includes the contribution of the other trivial carriers since PbTaSe$_2$ possesses several carrier pockets of trivial electrons and holes.
Thus, we need to extract them for a reasonable estimation of the anomalous Nernst effect.
Here, we assume that the contribution of other trivial carriers is proportional to the magnetic-field strength as observed in several materials \cite{11,12}.

\begin{figure}[t]
	\begin{center}
		\includegraphics[width=0.7\linewidth]{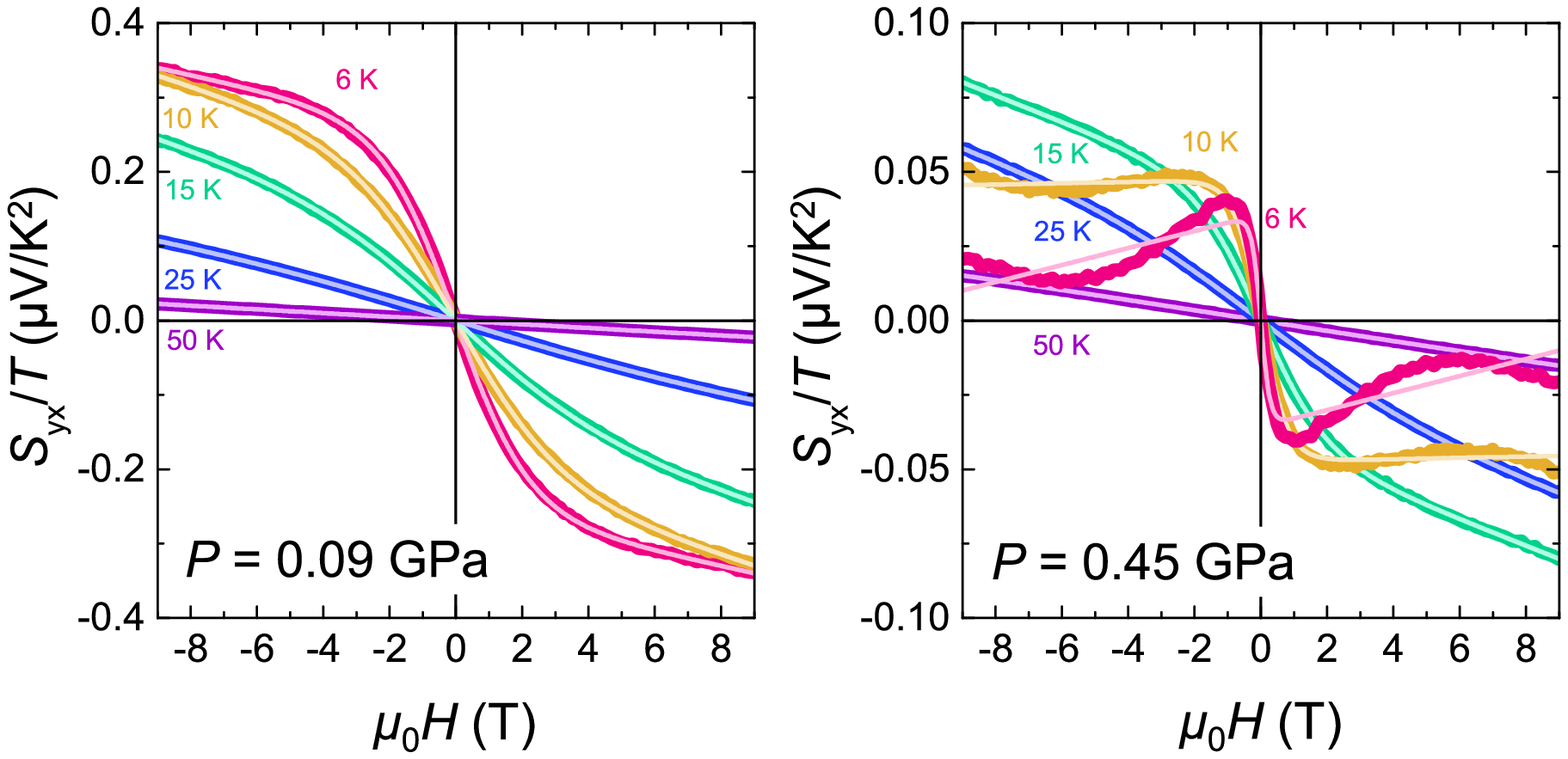}
		\caption[Structure]{\label{8}(Color online) Fitting analysis results for low-pressure (left panel) and high-pressure (right panel) phases at various temperatures, where bold curves and light-colored thin curves represent the experimental data and the simulation, respectively. Equation \ref{2} can also reproduce the experimental data in the other pressure and temperature (not shown here). }
	\end{center}
\end{figure}

Figure \ref{8} shows the fitting results in various temperatures. 
In the low-pressure phase, the experimental results are reproduced by Equation (1).
In the high-pressure phase, although the fitting curves slightly deviate from the experiment due to the local minimum around $5$ T, the magnetic field dependence of $S_{yx}$ is reproduced by Equation (1), and the amplitude of $S_{yx}^{ANE}$ is estimated qualitatively.

\begin{figure}[b]
	\begin{center}
		\includegraphics[width=0.65\linewidth]{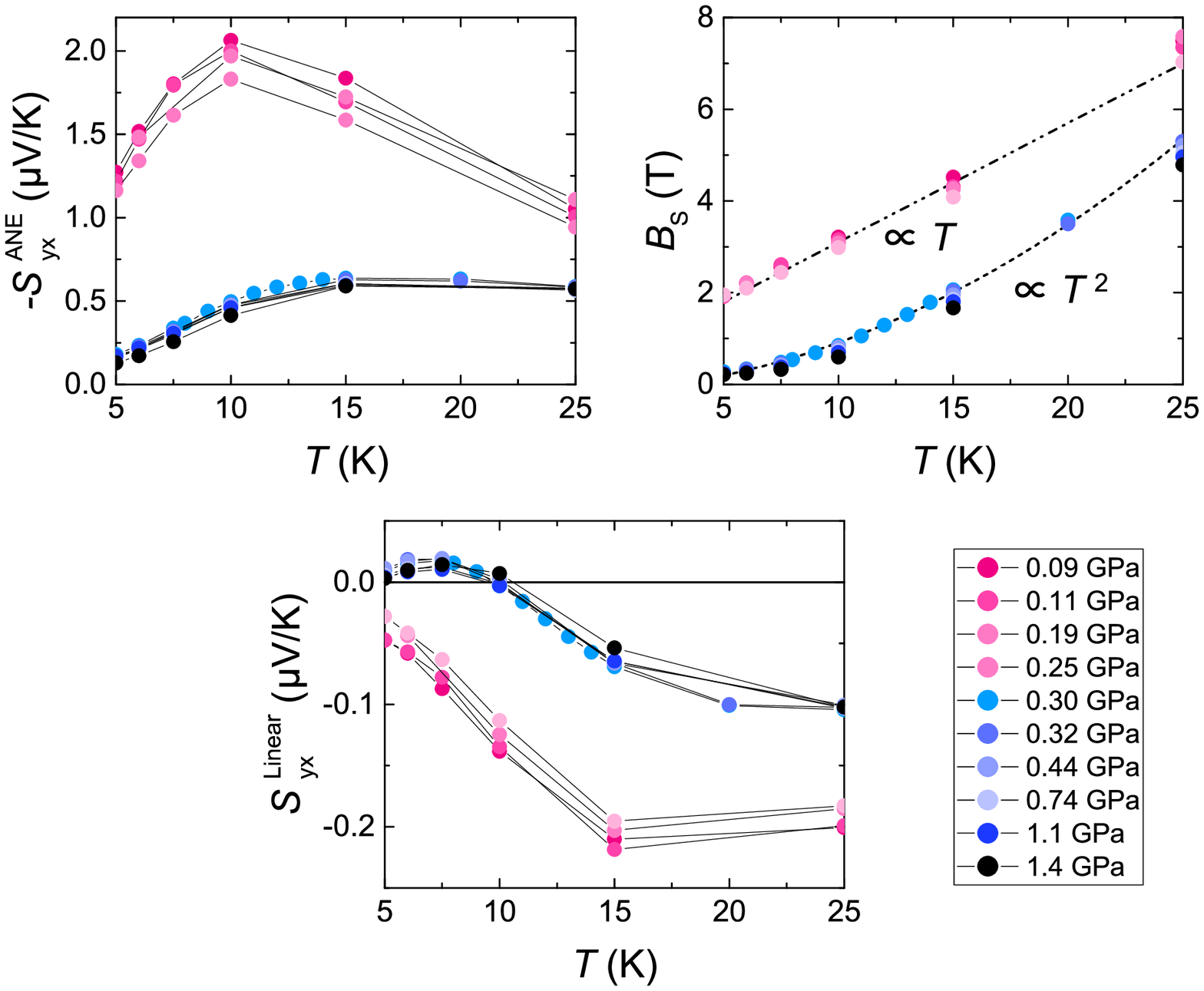}
		\caption[Structure]{\label{9}(Color online) Temperature dependence of fitting parameters: (a) anomalous term $S_{yx}^{ANE}$, (b) saturation field $B_s$, and (c) contribution of trivial carriers $S_{yx}^{Linear}$ at various pressures. }
	\end{center}
\end{figure}

Figure \ref{9} shows the temperature dependence of fitting parameters, $S_{yx}^{ANE}$, $B_s$, and $S_{yx}^{Linear}$ at various pressures.
After the structural transition, the peak temperature of $S_{yx}^{ANE}$ increases and the sign and behavior of $S_{yx}^{Linear}$ change.
The shift of peak temperature of $S_{yx}^{ANE}$ implies a shift in the band crossing point relative to the Fermi level \cite{13}.
The temperature dependence of $B_s$ becomes parabolic above the critical pressure.

\clearpage
\clearpage
\section{Magnetic field dependence of Seebeck coefficient}
A dip of the Seebeck coefficient near the zero fields is one of the characteristic thermoelectric features in topological semimetals \cite{7,8}.
This characteristic behavior can be reproduced by the following formula based on the Boltzmann model \cite{14}:
\begin{equation}\label{1}
	S_{xx} = \frac{\pi^2 k_{B}^2 T}{3e} \left(\frac{\sigma_{xx}^2}{\sigma_{xx}^2+\sigma_{xy}^2}\times D + \frac{\sigma_{xy}^2}{\sigma_{xx}^2+\sigma_{xy}^2}\times D_H\right),
\end{equation}
where $k_B$ is the Boltzman constant, $D$ and $D_H$ are fitting parameters that determine the Seebeck coefficient in zero and high field limits.
The longitudinal and Hall conductance $\sigma_{xx}$, $\sigma_{xy}$ are represented by the carrier mobility $\mu$ and carrier density $N_e$ as follows: $\sigma_{xx} = \frac{N_e e \mu }{1+(\mu B)^2}, \sigma_{xy} = \frac{N_e e \mu^2 B}{1+(\mu B)^2}$.
Substituting these formulas into Equation \ref{1} gives,
\begin{equation}\label{2}
	S_{xx} = \frac{\pi^2 k_{B}^2 T}{3e} \left(\frac{1}{1+(\mu B)^2}\times D + \frac{(\mu B)^2}{1+(\mu B)^2}\times D_H\right).
\end{equation}
In the case of trivial electrons described by the free-electron approximation, $D$ is equal to $D_H$ so that Equation \ref{2} does not depend on magnetic fields. 
Thus, the magnetic field dependence of $S_{xx}$ reproduced by Equation \ref{2} indirectly proves the existence of a nontrivial carrier, namely, Dirac/Weyl fermions.

\begin{figure}[b]
\begin{center}
	\includegraphics[width=1.0\linewidth]{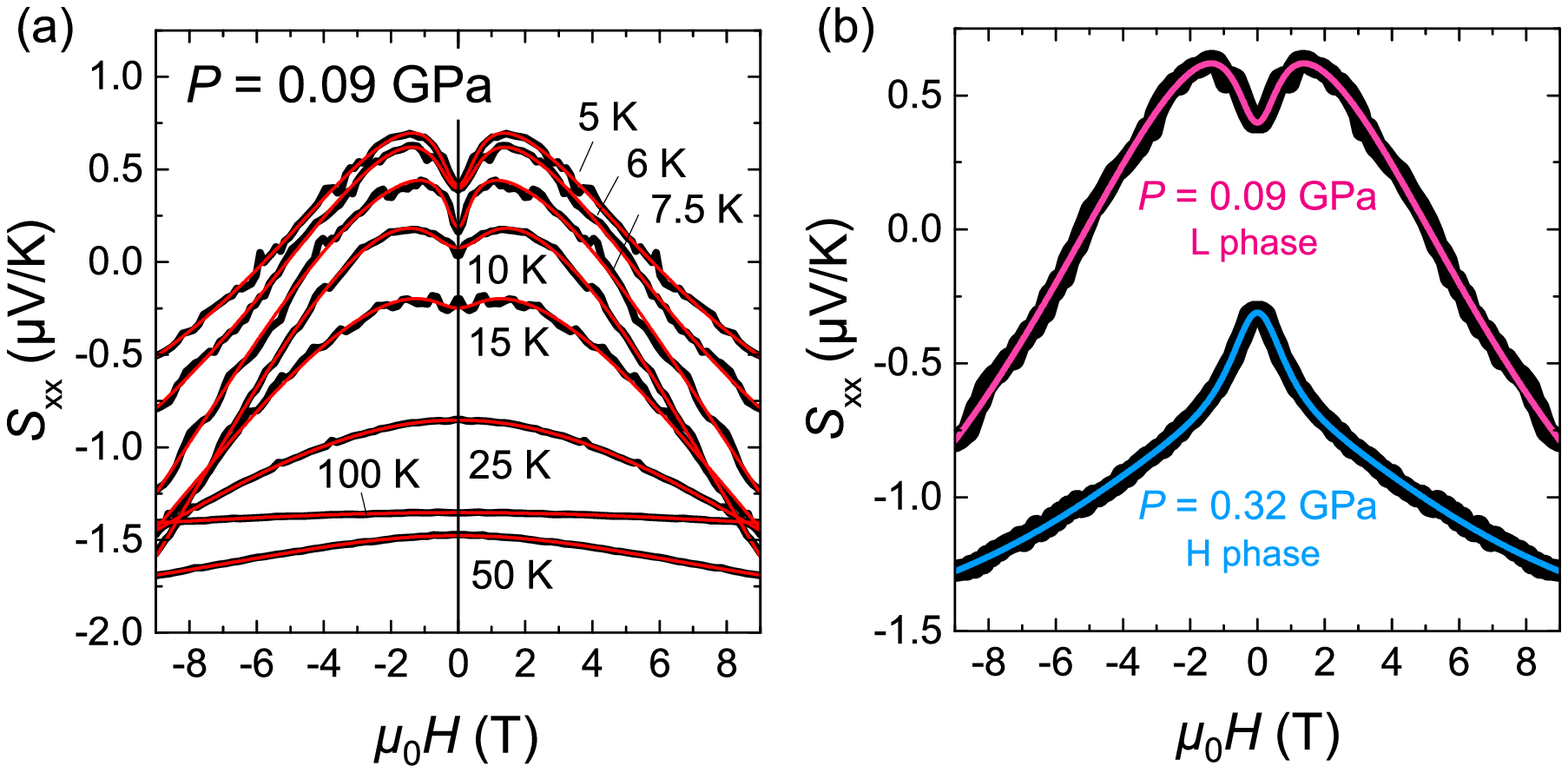}
	\caption[Structure]{\label{10}(Color online) Magnetic field dependence of $S_{xx}$ (bold black curve) and simulation curve (colored curve) in (a) low-pressure phase at various temperatures and (b) H and L phases at 6 K. Theoretical curves (Equation \ref{1} + Equation \ref{2}) reproduce experimental data well.}
\end{center}
\end{figure}

As shown in Figure \ref{10}, in PbTaSe$_2$, the magnetic field dependence of Seebeck coefficient in both low- and high-pressure phases could be reproduced by Equation \ref{1}, including the multi-band contribution \cite{15}:
\begin{equation}\label{sxxOrd}
	\frac{\sigma_{xx}^{ele}(B) \times S_{xx}^{ele} + \sigma_{xx}^{hole}(B) \times S_{xx}^{hole}}{\sigma_{xx}^{ele}(B)+\sigma_{xx}^{hole}(B)} \left(\sigma_{xx}^{i}(B) = \frac{\sigma^{i}_{0}}{1+(\mu^{i} B)^2}\right),
\end{equation}
where $\sigma_0^i$ is the conductivity at 0 T, $\mu^i$ is carrier mobility, and $S_{xx}^i$ is a constant parameter of the Seebeck coefficient.
We discovered that the parameters $D$ and $D_H$ change their signs in the structural transition, indicating the sign change of electrical charge of nodal line fermions.
Further theoretical investigations for the electronic state of PbTaSe$_2$ are needed to verify this scenario.

